\documentclass[]{spie}  
\pdfoutput=1 
\usepackage[]{graphicx}
\usepackage{subcaption}
\usepackage{cite}

\title{Developing adaptive secondary mirror concepts for the APF and W.M. Keck Observatory based on HVR technology}

 \author{Philip M. Hinz\supit{a},Rachel Bowens-Rubin\supit{a}, Christoph Baranec\supit{c}, Kevin Bundy\supit{a}, Mark Chun\supit{c},  Daren, Dillon\supit{a}, Brad Holden\supit{a}, Wouter Jonker\supit{b},  Molly Kosiarek\supit{a}, Renate Kupke\supit{a}, Stefan Kuiper\supit{b}, Olivier Lai\supit{f}, Jessica R. Lu\supit{e}, Matthew Maniscalco\supit{b}, Matthew Radovan\supit{a}, Sam Ragland\supit{d}, Stephanie Sallum\supit{g}, Andrew Skemer\supit{a}, Peter Wizinowich\supit{d}
\skiplinehalf
\supit{a} UC Santa Cruz, 1156 High St, Santa Cruz CA 95064, USA; \\
\supit{b} TNO Technical Sciences, Delft, The Netherlands; \\
\supit{c} Institute for Astronomy, University of Hawai`i at M\={a}noa, Hilo, HI 96720, USA \\
\supit{d} W. M. Keck Observatory, 65-1120 Mamalahoa Hwy., Kamuela, HI 96743, USA \\
\supit{e} UC Berkeley, 110 Sproul Hall Berkeley CA 94720, USA \\
\supit{f} Observatoire de la Côte d'Azur, Nice, France \\
\supit{g} UC Irvine, 510 E. Peltason Dr., Irvine CA 92697, USA \\
}
\authorinfo{Send correspondence to P. Hinz, E-mail: phinz@ucsc.edu}

\begin{document} 
\maketitle 
 
\begin{abstract}
An Adaptive secondary mirror (ASM) allows for the integration of adaptive optics (AO) into the telescope itself.   Adaptive secondary mirrors, based on hybrid variable reluctance (HVR) actuator technology, developed by TNO, provide a promising path to telescope-integrated AO. HVR actuators have the advantage of allowing mirrors that are stiffer, more power efficient, and potentially less complex than similar, voice-coil based ASM’s.  We are exploring the application of this technology via a laboratory testbed that will validate the technical approach.  In parallel, we are developing conceptual designs for ASMs at several telescopes including the Automated Planet Finder Telescope (APF) and for Keck Observatory. An ASM for APF has the potential to double the light through the slit for radial velocity measurements, and dramatically improved the image stability.  An ASM for WMKO enables ground layer AO correction and lower background infrared AO observations, and provides for more flexible deployment of instruments via the ability to adjust the location of the Cassegrain focus.   
\end{abstract}

 
\section{Introduction}
Adaptive optics (AO) is an important capability of ground-based telescopes that enable higher spatial resolution, more sensitive observations.  Integrating the AO into the telescope optics (via an adaptive secondary mirror) can significantly improve sensitivity for both diffraction-limited AO (especially at 2-5 $\mu$m), and seeing-limited instruments that employ ground-layer adaptive optics (GLAO).  Current generation adaptive secondary mirrors (ASM’s) have been implemented at several telescopes, including the MMT, LBT\cite{Riccardi2010}, Magellan,  and VLT using voice coil actuators and thin glass shells to replace the standard secondaries (see \cite{Biasi2010} for an overview).   While these devices are in routine operation, several aspects make them non-ideal, tied to the low power efficiency of current actuator designs.  

TNO (The Netherlands Organization) has developed a technology with efficiencies approximately 40 times that of current actuators\cite{Kuiper2018}. The actuators utilize the efficiency gains of enclosing the magnetic field path that moves the actuator in a ferromagnetic material, thus reducing the current needed to apply a particular force.  This larger force allows for building in an internal stiffness to each actuator and rigid connections to the facesheet.  The resulting assembly has very high structural resonant frequencies, compared to voice coil designs, allowing a simple control approach.  Further, the power required to correct turbulent wavefronts can be dramatically lower, allowing for simpler, passive cooling approaches to the system.  Finally, the additional efficiency can be traded against the facesheet thickness to provide a sturdier deformable membrane that reduces fabrication risk and complexity.  The resulting system has the potential to be more robust and simpler than currently deployed ASM's. 

Development of this technology can be aided by lab and on-sky devices that are used on small and medium aperture telescopes, demonstrating the readiness of the technology for large aperture telescopes and next generation observatories such as the TMT.  Below, we review some of the motivations for an ASM-based AO system for large aperture telescopes like W.M Keck Observatory (WMKO) and next generation of extremely large telescopes.  We describe concepts for a lab-based ASM prototype, called FLASH, a concept for an ASM on the 2.4 m Automated Planet Finder (APF) telescope, and the current status of a concept study for the Keck telescopes.

\section{Motivation for ASM-based AO}

The value of integrating an ASM into a telescope facility primarily lies in the potential for broader use of AO by any instrument that can benefit from the improved image quality.  We review below two areas where an ASM-based AO system can be advantageous.

\subsection{Improved Thermal Infrared Observations}

AO systems that are designed for diffraction-limited operation with post-focal plane optics require additional optical surfaces to form an image of the pupil and provide field rotation compensation. Taking the Keck AO bench as an example, the system introduces 7 reflections after the tertiary mirror for a total of 10 reflections. A dichroic also splits the light between science instrument and wavefront sensor (WFS).  We assume the primary mirror and tertiary mirrors are 95\% reflective in the infrared (up-facing mirrors tend to not stay as clean as the down-facing mirrors) and the secondary and each post-focal mirror is 97.5\% reflective. We further assume the dichroic is 96\% transmissive.  The resulting throughput of the system is 70\%.  The emissivity (one minus the throughput) of the system is 30\%. 

An ASM-based system would only have the telescope optics contributing emissivity at the level of 12\%. The throughput would be improved by eliminating four reflections to 78\%. The field rotator and dichroic could be cooled to eliminat their contribution to any emissive background.  

Similar to the calculations shown in \cite{Lloyd-Hart2000}, the background for these two comparable systems can be calculated using blackbody emission of the typical nighttime temperature of the telescope over the passband of interest with the emissivity, $em$, for each system as calculated above.  The resulting background flux in photons/s/m$^{2}$/nm/arcsec$^{2}$ is given by  

\begin{equation}
    F_{tel} =  \frac{2\cdot c \cdot \pi^{2}}{(180 \cdot 3600)^{2} \lambda^4}\frac{em \cdot 10^{-9}}{e^{\frac{h\cdot c} { \lambda \cdot k \cdot T_{tel}}} - 1  )}
\end{equation}

where c, h,  and k are the speed of light, Planck's constant, and Boltzmann's constant respectively, and T$_{tel}$ is the temperature of the telescope optics.   Figure 1 shows the resulting background at two relevant passbands, K and L.  Plotted along with these is an estimate of the sky brightness (F$_{Sky}$) from Gemini\cite{Gillett1998} for Maunakea at an airmass of 1.5 and PWV of 1.6 mm.

One notices that there are regions of the K and L band where the telescope emission dominates over the sky background.  For these regions of the spectrum, an ASM-based system can provide the most significant improvement in sensitivity.  It is possible to estimate the relative time required to achieve a particular sensitivity\cite{Lloyd-Hart2000}, if the noise is dominated by the effects described here (as opposed to, say, read-noise from the camera detector, or photon noise from the source itself).  

\begin{equation}
    \frac {t_{postfocal-AO}} {t_{ASM-AO}} = \left(\frac{Thpt_{ASM-AO}}{Thpt_{postfocal-AO}}\right)^2 \frac{F_{tel-postfocal-AO}+Thpt_{postfocal-AO} \cdot F_{sky} }{F_{tel-ASM-AO}+Thpt_{ASM-AO} \cdot F_{sky}}
\end{equation}

where $Thpt$ are the throughput numbers for each system above. The right panel of Figure 1 shows this relative time.  Time improvements of factors of 2 to 3 are possible in regions of the K and L photometric bands.

   \begin{figure}
   \begin{center}
   \begin{tabular}{c}
   \includegraphics[height=14cm]{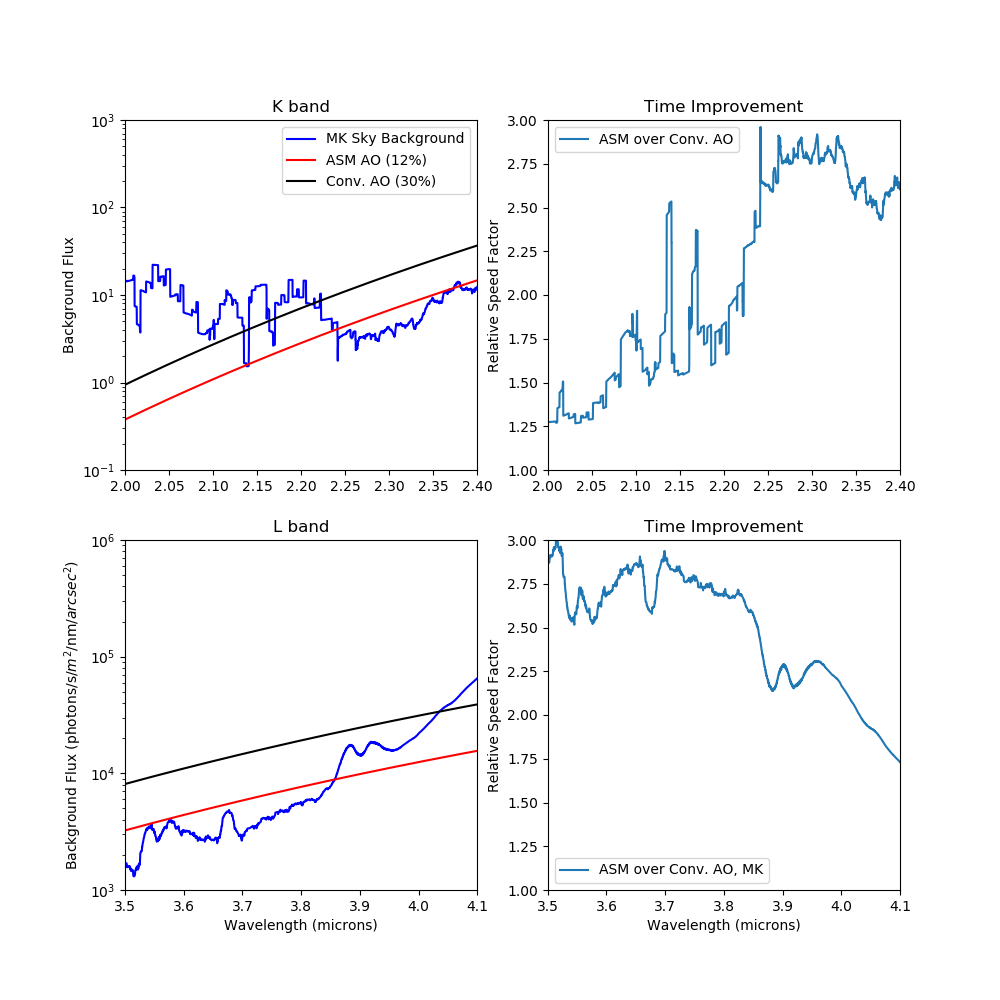}
   \end{tabular}
   \end{center}
   \caption[example] 
   { \label{fig:example} 
Calculation of the background telescope emission for a post-focal plane AO system, compared to an ASM-based AO system at K and L bands. The modeled sky emission for Maunakea at a PWV of 1.6 mm and airmass 1.5 from Gemini is also included.  The right panels show the effect on time to reach a particular SNR when comparing these two systems.}
   \end{figure}

\subsection{Ground Layer Correction}

Studies of many ground-based astronomical sites show that a large fraction of the optical turbulence arises within the first few hundred meters of the ground (including Mt. Graham\cite{Masciadri2010}, San Pedro Martir\cite{Avila2004}, Paranal\cite{Osborn2018}, and Antartica\cite{Hagelin2008}).   AO systems that correct for only the ground layer of turbulence, ground-layer adaptive optics system (GLAO) , take this fact into consideration and provide a partially corrected wavefront over fields of view several arcminutes in size.  Facility GLAO systems have been deployed at several telescopes, including the LBT \cite{Rabien2019} and the VLT.   In the case of Maunakea, Hawaii it appears that the ground-layer is dominated by a very thin surface layer with most of the optical turbulence within the first ~30 meters \cite{Chun2009}\cite{Chun2014},\cite{Schok2009},  .  GLAO over fields of view of 18 arcmin has been demonstrated on Maunakea\cite{Chun2018}.

An ASM enables these very wide fields of view since the design/fabrication of the optical relay system needed for a conventional deformable mirror system is difficult.  The optical designs are driven to large optics (reducing the physical angles on the elements in the relay) and aspheric surfaces\cite{Chun2010}.  We note that the image quality and pupil quality over the field must meet strict optical requirements.  By making the wavefront correction within the telescope optics, we avoid the latter problem and a simple field correction (e.g. a field flattener lens) allow these large corrected fields to be obtained.   The Subaru\cite{Rigaut2018}, and Keck\cite{Lu2018} telescopes on Maunakea are developing or are studying GLAO systems for their telescopes.   Performance simulations\cite{Lu2018} have been developed to show that a GLAO system on Keck will improve the FWHM by 1.5x for visible (500 nm) and 2x for NIR (1.5 $\mu$m) observations.   The Gemini Observatory studied this option in the past\cite{Szeto2006} and are reconsidering these gains given the maturing and developing ASM technologies.  An ASM based on the TNO actuators\cite{Kuiper2020} is being developed\cite{Jonker2020} for UH2.2-meter telescope\cite{Chun2020}. 

\section{Testing TNO actuator technology}

TNO has built several mirrors with actuators on an 18 mm spacing that have been used in AO demonstrations and test setups. One such device (dubbed DM3 by TNO) has been characterized in the Lab for Adaptive Optics (LAO) at UCSC\cite{Bowens-Rubin2020}. Measurements of the stroke, influence functions, linearity and hysteresis have been carried out, verifying the suitability of these devices for use in large format deformable mirrors.

The first on-sky ASM using this technology will be deployed to, and demonstrated at, the University of Hawaii 2.2-meter telescope (UH88) on Maunakea, HI\cite{Chun2020}. To support the deployment of this TNO ASM at the UH88, we have worked with TNO to build the First LAO Adaptive Secondary Holophote (FLASH).  This is a 19 actuator device with the same actuator spacing and facesheet thickness as the UH88 ASM. The device has an aperture of 160 mm. The facesheet has been manufactured and coated from standard polished borofloat sheets that have been pressed flat.  The slumping technique explored for FLASH is anticipated to be the standard technique for generating future facesheets for the APF and Keck ASM.  FLASH is due to be delivered in late 2020, and will be characterized in a similar way to the DM3.  In addition we anticipate carrying out cold and gravity tests of the device as well as capacitive sensor measurements of both its high speed response and long term stability. 

\section{MOTIVATION FOR AN ASM FOR APF}
Adaptive optics has the potential to improve both the observational efficiency (by concentrating the point spread function of the telescope) and the instrumental stability (by creating a PSF that is more stable than is possible with a seeing-limited telescope) of APF.   The approach of designing an AO system for the APF is to deploy a system capable of improving the image quality, while minimizing the changes to the spectrometer and calibration system.  Thus, we have elected to build the AO system into the telescope itself. 

Instruments such as the APF/Levy tend to observe bright stars in order to make precise radial velocity (RV) measurements.  As such, there are plenty of photons available to measure and correct the wavefront of the light going into the spectrometer.  Further, there is no anisoplanatism effect since the “guide star” used for wavefront control is the science object itself. This allows for a straightforward and simple implementation of the adaptive optics system. One requires only an on-axis, natural guide star, wavefront sensor.

 For typical observations, the APF uses an image plane slit width of 0.5 arcsec.  Seeing on Mt. Hamilton averages 1.5 arcsec.  Analysis of guider data for a typical sequence allows us to estimate the slit throughput as ~21\% with variations of 4\% as seen by the guider over 1 second integrations (see Figure 2 left).  Some of this throughput is due to atmospheric seeing.  However, we see a significant portion of the throughput reduction to be due to slow ($<$1 Hz) tracking variations.

   \begin{figure}
   \begin{center}
   \begin{tabular}{c}
   \includegraphics[height=6cm]{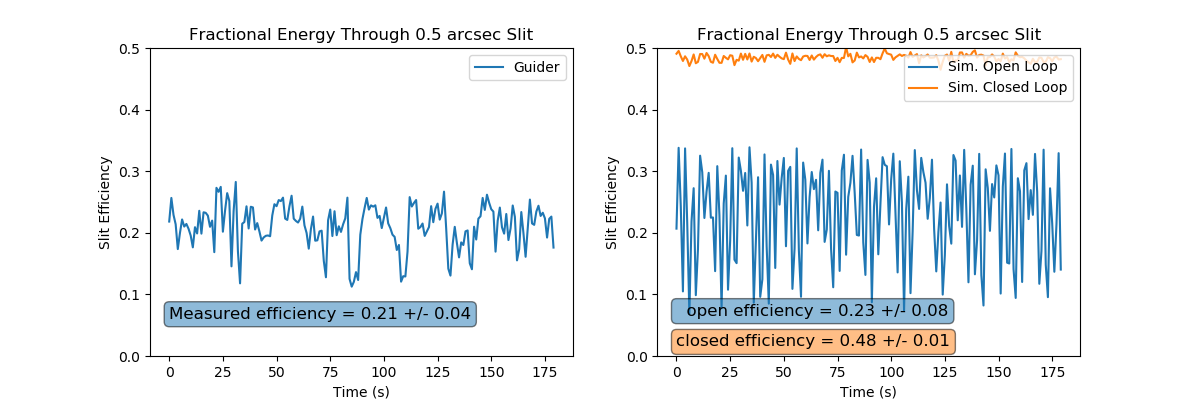}
   \end{tabular}
   \end{center}
   \caption[example] 
   { \label{fig:example} 
Current APF measured slit efficiency as measured using the guider (left) and simulated efficiency (right) without and with AO correction. 
}
   \end{figure}

 We can simulate the expected improvement.  We have used the python package HCIpy\cite{Por2018} to create an atmosphere with similar seeing to the guiding data and have added in slow tracking variations at 0.3 and 0.05 Hz.   The resulting slit efficiency (23\%) and variations of 8\%  are similar to what is seen by the APF guider (Figure 2, right).  By carrying out closed loop AO corrections we can increase the efficiency by 2.1 times.  An additional benefit is the variation in efficiency is reduced to about 1\%. For the 1 arcsec slit width the improvement is similar: without AO correction the efficiency is 45\% while after correction it is 72\% for a 1.6 improvement factor.

The connection between PSF and RV stability is complex. One example of their correlation is seen in HARPS data (admittedly, a different style spectrograph to APF/Levy) where Lo Curto  et al.\cite{LoCurto2010} note that a guider variation of 0.1 arcsec results in approximately 0.3 m/s errors.
Figure 3 shows the improved stability, compared to current performance an AO system can provide.  Offset variations, seen at the APF with a standard deviation of ~0.2 arcsec at 1 Hz (Figure 3 left top and right top simulated open loop data)  can be reduced to 0.004 arcsec (Figure 3 right simulated closed loop data).  Similarly the full width at half maximum (FWHM) estimates for 1 s exposures are reduced from 1.2-1.3 arcsec. to 0.05 arcsec.  The variations in the FWHM are reduced from 0.1-0.2 arcsec. to 0.001 arcsec.  In other words, the point spread function variations in the resulting spectrum can be reduced by dramatic amounts.   While it is not straightforward to translate this image stability improvement into RV stability improvement, this setup will allow us to test what the value is for AO-integrated RV spectrometers.

   \begin{figure}
   \begin{center}
   \begin{tabular}{c}
   \includegraphics[height=12cm]{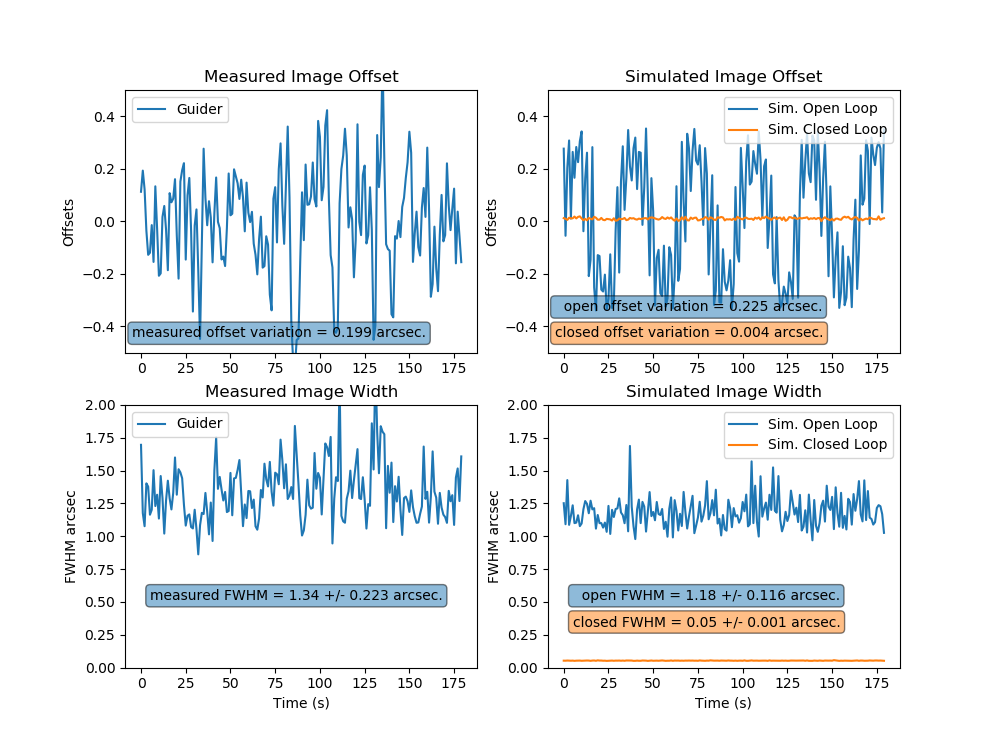}
   \end{tabular}
   \end{center}
   \caption[example] 
   { \label{fig:example} 
Current APF image offset (top left ) and image width (bottom left) compared to  simulated image offset (top right) and image width (bottom right) for open and closed loop adaptive optics correction.  The AO will both reduce the image offset variations and improve the image width and stability of the PSF variations. 
}
   \end{figure} 

How can we determine whether the image motion and PSF variations affect the quality of the RV measurements? As a comparison, during the nights of Aug 22 and 23 in 2016, two stars were observed simultaneously with the APF/Levy spectrometer and the Planet Finding Spectrograph\cite{Crane2010} (PFS) on the Magellan Clay Telescope. The observations were made so to ensure that both data sets had the same start time for each exposure. Both instruments have an Iodine cell; both are mounted on the telescope at the Nasmyth; both use apertures that depend on the natural seeing; both have the same optical design, and the data were reduced by the same individual, Paul Butler, using the same software.
 
The main differences between the APF/Levy and Magellan/PFS is that the latter has both a larger aperture and better seeing. For both sets of observations of the same star, an additional source of error, beyond the shot-noise measurement errors, is required to explain the scatter in the data. For the APF/Levy, that additional error is 1.2 m/s, as compared with only a 0.6 m/s additional scatter from PFS, see Figure 4. This additional scatter of 1.2 m/s matches the floor in the measurement error found by Burt et al.\cite{Burt2014} for quiet, bright stars on the timescale of months. Since the main difference between the instruments is the better seeing for the PFS, it is plausible that the typical additional scatter seen at APF is due to image motion on the spectrometer slit.

Even if only the efficiency improvement can be realized, the value of AO for the APF is to increase the cadence of nightly observations.  The efficiency improvement suggests that we could be able to observe essentially double the number of stars per night compared to current observations.  We expect to provide this improvement on stars as faint as I=11. This dramatic speed up of observations is equivalent to having a second APF available.

  \begin{figure}
   \begin{center}
   \begin{tabular}{c}
   \includegraphics[height=6cm]{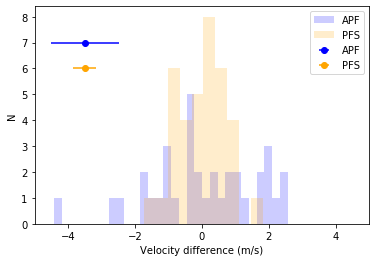}
   \end{tabular}
   \end{center}
   \caption[example] 
   { \label{fig:example} 
Histogram of RV scatter seen by the APF (blue) compared to PFS (orange) for simultaneous observations of HD 3651. The typical measurement uncertainty for each instrument is shown in the upper left.  Since the two instruments have nearly identical optics and reduction software, an additional factor appears to be affecting the APF, resulting in over double the internal scatter (1.3 m/s) compared to the PFS data (0.6m/s).   It is plausible that this is due to the worse image motion and seeing for the APF.  
}
   \end{figure}

\section{AO for APF Concept}

The Automated Planet Finder (APF) is a 2.4 m telescope with a high spectral resolution iodine cell-referenced spectrometer\cite{Vogt2014}. Typical targets are relatively bright stars that can be monitored for small radial velocity variations due to orbiting planets. Because of this targeted mission, the design of an AO system to stabilize the PSF for the APF can be relatively straightforward.  We require a wavefront sensor that uses light not employed for science observations. A suitable passband is 700-1000 nm.  Since the spectrometer is observing a bright star, the WFS will be fixed on-axis.  We set the spatial sampling at 8 subapertures across the telescope diameter. This will set both the deformable mirror actuator geometry and Shack-Hartmann lenslet size.  We arrived at this value after exploring different spatial samplings in the AO simulation.  While better samplings improved the Strehl, the coarse slit size resulted in little improvement in throughput above this value.

\subsection{ASM}

The deformable mirror will be the 37 cm diameter telescope secondary mirror.  The high efficiency HVR actuators will allow for us to deploy a secondary mirror without the need for any active cooling of the assembly. This allows for a very compact and relatively simple assembly as shown in Figure 5, the ASM will replace the fixed secondary mirror, while allowing for reuse of the focus and tip/tilt mechanism built into the APF top-end structure. 
The notional actuator geometry is shown in Figure 5 (right).  The actuator spacing is similar to the UH-88 design and to the FLASH prototype being tested in the Laboratory for Adaptive Optics. 
To match the existing APF design, the facesheet will be a hyperbola shape with a radius of R=1198 mm and a conic constant of -1.49. The facesheet can be generated using a slumping of borofloat flat polished sheets. Tests are underway in the LAO to validate this fabrication approach.

\begin{figure}[h!]
\centering
\begin{subfigure}{0.40\textwidth}
  \centering
  \includegraphics[width=1.0\linewidth]{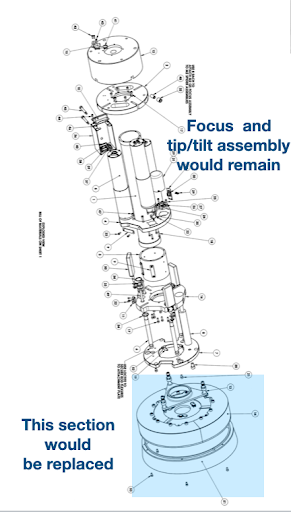}
\end{subfigure}
\begin{subfigure}{0.40\textwidth}
  \centering
   \includegraphics[width=1.0\linewidth]{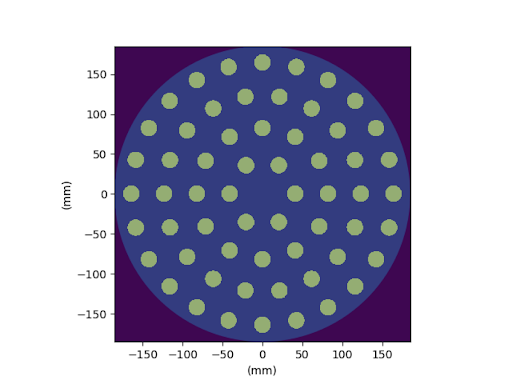}
 \end{subfigure}
 \caption[example]{The APF secondary assembly (left) and  layout of the ASM actuators (right).  Three parallel screw assemblies provide focus tip and tilt of the secondary assembly.  These will be reused for the ASM upgrade. 
 }
   \end{figure}

\subsection{WFS}
Since the wavefront sensor requires good sensitivity in the red, and good noise performance to be able to operate on the faintest stars for an APF observation, electron multiplied CCD (EMCCD) devices are ideal.  Alpao has developed a Shack-Hartmann (SH) EMCCD device with low latency and high framerates suitable for our needs.  The device has 16x16 sub-apertures and can operate at up to 1004 Hz.  Since our needs are to have 8x8 SH, we can operate the device at up to 1838 Hz with a frame latency of 69 $\mu s$.  
The SH wavefront sensor will use the light from 700-1000 nm, as shown in Figure 6. The average quantum efficiency over this region is around 50\%.  Assuming a 30\% (not including the detector) system throughput, and using the specified noise for each subaperture from Alpao we estimate being able to provide usable stabilization down to guide star magnitudes corresponding to an I band magnitude of approximately I=11.  Simulations with the HCIpy software package\cite{Por2018} confirm this, indicating 1 kHz operation will be optimal to stars as faint as I=9, 300 Hz should be used for stars with 10-12, and that some improvement is possible using 100 Hz framerate down to I=14 (see Figure 6, right). 

The WFS will be fed by a short pass dichroic that has a cut-on wavelength of approximately 700-750 nm (to be determined in collaboration with APF observers).  The dichroic will feed a transmissive reimaging optic that images the APF pupil onto the Alpao Shack-Hartmann lenslet array. The fore optics for the Levy spectrometer are shown in Figure 7.  They are laid out on an optical breadboard that has suitable space for a dichroic, reimaging lens and Alpao WFS module.

\begin{figure}[h!]
\centering
\begin{subfigure}{0.40\textwidth}
  \centering
  \includegraphics[width=1.0\linewidth]{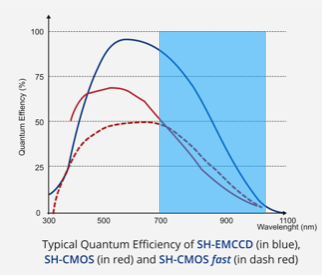}
\end{subfigure}
\begin{subfigure}{0.40\textwidth}
  \centering
   \includegraphics[width=1.0\linewidth]{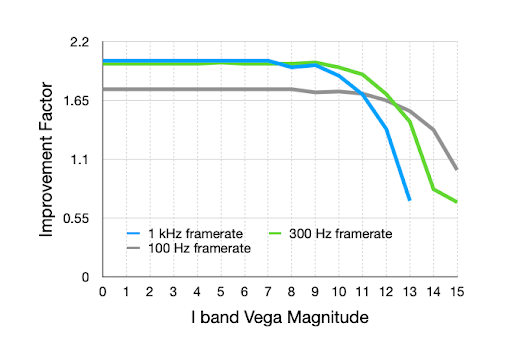}
 \end{subfigure}
 \caption[example]{Quantum efficiency of the Alpao WFS (left) from the EMCCD factsheet (see alpao.com) and the simulated Efficiency Improvement  vs. Magnitude with AO compared to no correction. 
 
 }
   \end{figure}

\subsection{APF AO system}
Currently the APF operates with a guider camera providing pointing information to the telescope at a 1 Hz rate.  This system will be retained and used for an acquisition camera for the system.  Once the star is acquired on the WFS, the guiding corrections will be turned off and the AO loop will provide stabilization and, if needed, pointing offsets for the telescope tracking software.  
The real time computer will be based on the Keck RTC setup developed for the infrared pyramid WFS on Keck.  As such, the software implementation involves an adaptation of an existing system, rather than implementation of any new AO software.  

  \begin{figure}
   \begin{center}
   \begin{tabular}{c}
   \includegraphics[height=6cm]{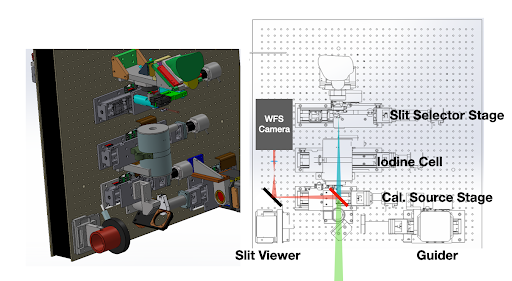}
   \end{tabular}
   \end{center}
   \caption[example] 
   { \label{fig:example} 
 Model of the APF optics, showing the location for a WFS camera. Light arrives from the bottom and is directed to the guider on the bottom right.  The remaining light goes through several stages that allow for optics to be inserted in the beam.  We will use the calibration source stage to allow insertion of a short pass dichroic in the beam.  The dichroic will reflect long wavelength light to an Alpao SH WFS. A simple reimaging lens is all that is needed to create a pupil image on the SH WFS camera. 
}
   \end{figure}

\subsection{Calibration}

Once delivered, the ASM will be characterized via a deflectometry test.   The system will be setup to be illuminated by a large LCD screen.  The ASM will undergo a poke test to measure the influence functions of all the actuators.  Once these are measured, we will have an initial interaction matrix to deform the facesheet.  Further refinements of the interaction matrix will be achieved by applying modal shapes to the mirror, to map out achieved versus commanded surface deformations.
Once the deflectometry tests have been completed under standard laboratory conditions we will carry out the same measurements over a range of elevation angles from zenith to horizon.  UCO has a cold chamber that can carry out the same tests at temperatures down to -10 C .

  \begin{figure}
   \begin{center}
   \begin{tabular}{c}
   \includegraphics[height=6cm]{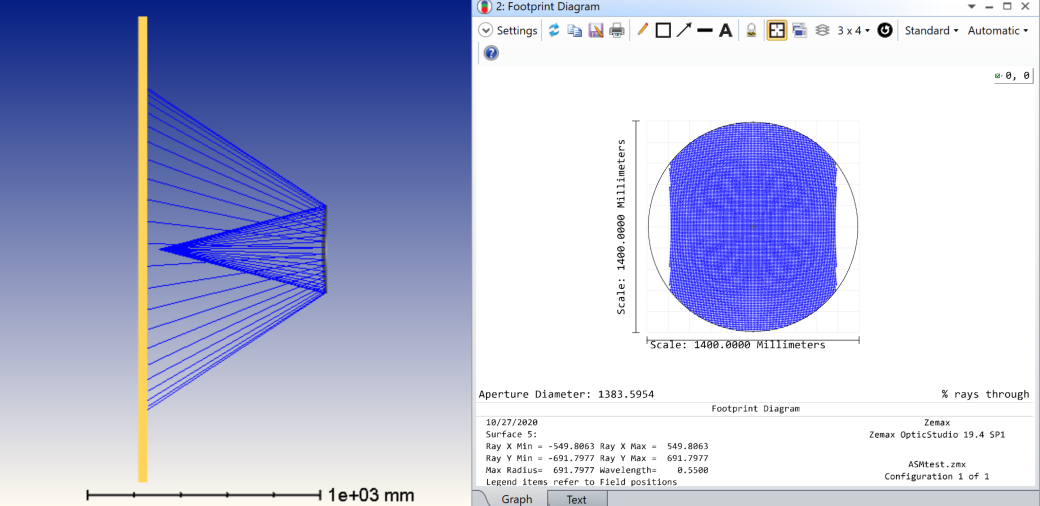}
   \end{tabular}
   \end{center}
   \caption[example] 
   { \label{fig:example} 
 Layout for a deflectometry test for the APF ASM.  A large (1.1 m x 2 m) LCD screen will be used to illuminate the ASM with a pattern that will be recorded by a camera.  The screen does not completely illuminate the optic as seen by the camera (right), but can be rotated to measure the complete surface. 
 
}
   \end{figure} 

\section {KECK ASM CONCEPT DEVELOPMENT}

The lab testing and APF ASM described above, along with the HVR ASM on the UH88 will provide technology platforms to demonstrate the suitability of this technology for use in astronomical facilities.  In parallel we are developing a concept for an ASM that can be deployed at Keck observatory. Development of the concept allows us to understand the scope and resources needed to deploy an ASM at WMKO.   

\subsection{Instrument Readiness}
An ASM facility promises to have broad application for AO-improved performance.  However, integrating such a different AO approach into existing and proposed instrumentation will require careful planning and identifications of issues early on in the project to understand what is possible. The work described here builds on the science case and instrument readiness studies lead by J. Lu and summarized in \cite{Lu2018}. While the work for the current study is still to be done, we summarize below the existing and future instruments likely to be used with an ASM-based AO system. 

\subsubsection{Existing Instruments}

 MOSFIRE, an infrared imager and multi-slit spectrograph at the Keck I Cassegrain focus, is perhaps best suited for a GLAO feed, although it lacks an atmospheric dispersion compensator. Differential image refraction separates images by about 0.3 arcsec at 2 airmass between 1 and 2.4 $\mu$m, so a lack of an ADC is limiting but GLAO can still be beneficial.  A replacement detector with smaller pixels than the current 0.18 arcsec. would be needed to take advantage of the GLAO image quality. This is possible with new generation H2RG devices from Teledyne. There is a feasible design for deploying laser-guide star wave-front sensors in front of MOSFIRE in the tertiary tower in order to measure the wave-front error
 
 LRIS and DEIMOS may be able to accept a GLAO feed based on their internal image quality; but, they lack either an atmospheric dispersion compensator (DEIMOS) or flexure compensation (LRIS) to utilize narrower slits matched to the GLAO image quality. 
 
 Diffraction-limited instruments such as OSIRIS on Keck I and NIRC2 and NIRSPEC+AO on Keck II make use of the AO bench for wavefront sensing and correction\cite{Wizinowich2006}. A replacement for the AO bench that selects light for wavefront sensing without additional optics would be needed to realize the improvements described above for an ASM-based AO system.  To replicate the AO bench functionality with these existing instruments, such a system would need to accomplish field rotation, and natural guide star and laser guide star wavefront sensing, as well as acquisition and tip-tilt sensing.  
 
 KCWI, a multi-object optical spectrograph with 0.15 arcsec. pixels and a FOV as large as 20x33 arcsec could potentially benefit from GLAO.  
 
 Instruments such as HIRES may benefit from the additional image stability an ASM could supply if a natural guide star wavefront sensor could be added to their guiding capabilities. 

\subsubsection{Future Instruments}

Several planned or proposed instruments could benefit from an ASM.  

An upgrade to the LRIS (LRIS2) is being planned that will incorporate a flexure compensation system and allow for reconfiguring the system to give space for a WFS package.  

The Santa Cruz Array of Lenslets for Exoplanet Spectroscopy (SCALES) is an infrared integral field spectrograph that works at wavelengths out to 5 $\mu$m where the lower background of an ASM would benefit sensitivity as well as AO performance.  

The Fiber Optic Broad-band Optical Spectrometer (FOBOS) is a proposed wide field (20’ diameter), moderate spectral resolution facility instrument for Keck. GLAO with an adaptive secondary would improve photometric depth achieved on single fiber collectors, enable crowded source targeting and create new scientific opportunities with critically sampled IFUs which spatially resolve galaxies beyond z $\sim$ 0.5.  

The Immersion Grating Near Infrared Spectrograph (IGNIS) is a high spectral resolution infrared spectrometer potentially providing improved performance relative to the existing NIRSPEC instrument on Keck, based on immersion grating spectrometers similar to IGRINS\cite{Park2014}.  Integrating such an instrument with the ASM has the potential to provide improved sensitivity at wavelengths beyond 2 $\mu$m.   

\subsection {Keck ASM Study Status}

The Keck ASM study is currently underway to develop a concept based on HVR technology with the following goals:

\begin{itemize}
    \item Develop a concept of an ASM suitable for integration with the Keck telescopes
    \item Develop a dynamical and thermal model for the design
    \item Define interfaces for integrating the device into the telescope top-end
    \item Develop an optical testing strategy for deploying and maintaining the ASM
    \item Develop a rough cost estimate
\end{itemize}

The intention is to develop the concept to the level that the scope of such an implementation can be understood, positioning the WMKO community to determine a suitable funding approach for future implementation. The study began in August 2020, with a plan to complete this stage of the effort in summer 2021. 

\subsection{Actuator Tradeoff}

The choice of the number of actuators for the study is driven by the scientific motivation of providing a device that can provide good GLAO and diffraction-limited AO performance for future generations of WMKO instruments.  HVR devices have been demonstrated on spacings as small as 18 mm and up to 40 mm.  For the Keck secondary diameter (1.4 m) this would result in total actuator counts from about 1000 total actuators up to about 4000 actuators. 

In the initial portion of this study, we considered three separate general concepts for this study: 1000, 2000, and 4000 total actuators (see Table 1 for a listing of parameters for each concept).  The actuators are assumed to be laid out in a circular grid for each concept with an additional outer ring of actuators for each concept, used to control the edge effects of the shell.  The requirement on the interactuator stroke drives the thickness of the shell specified for different actuator count ASM designs.  We assume in Table 1 that the interactuator stroke required scales with the actuator spacing.   The achievable deflection of an actuator also scales with the thickness of the facesheet cubed.  Taking these two effects into account we can estimate the thickness required for different actuator counts starting from the calculation that the 1000 actuator count achieves an interactuator stroke of 4 $\mu$m for a projected spacing of 350 mm. 

The 1000 actuator concept would already provide improved AO performance compared to the current Keck AO bench (which has a fitting error of 190 nm using the same atmospheric parameters as in Table 1).  Still, we expect that instruments such as SCALES that depend on high contrast AO performance for achieving their science goals would be limited by the fitting error introduced by this low number of actuators. Similarly GLAO at shorter wavelength could benefit from a finer actuator pitch. For 4000 actuators there are two effects that may make this design less desirable.  First, the smaller separation drives the facesheet to smaller thicknesses, to achieve similar levels of interactuator stroke.  This potentially makes the shell more difficult to manufacture and possibly more fragile to handle.  In addition, the weight for the full system becomes significantly more for this large actuator count.  Because of these competing technical issues we have decided to develop further the concept for 2000 actuators for this study.  It is useful to note that these alternative concepts are all viable should the motivations for deploying an ASM at WMKO be different from those considered for this study.

\begin{table}[h!]
\center
\caption{\textbf{ Parameters for a Keck ASM for different total actuator count}}
\begin{tabular}{|l|r|r|r|}
\hline
\textbf{Target Number of Actuators} & \textbf{1000}     & \textbf{2000}     & \textbf{4000}     \\ \hline
Actuator spacing on ASM (mm)            &  36           &       26          &        19            \\
Projected spacing on Primary (mm)       &  346          &       245         &       173               \\
Actuators Rings                         &  19           &        26         &        37               \\
Actuators in Clear Aperture              & 1027          &       1951        &       3997            \\
Total Actuators                         & 1135          &   1951            &       4213            \\
Fitting Error for r$_{0}$= 10 cm (nm)   & 123            &       92        &         69            \\ \hline
Facesheet thickness (mm)                & 3.3            &     2.9          &        2.4    \\ 
Moving Mass                             &   432         &    577            &       855     \\
Total Mass                              &  982          &   1127            &   1405        \\ \hline
\end{tabular}               \end{table}

\subsection{Current TNO Concept}

Figure 9 shows the current concept for an HVR ASM at WMKO.   The device could be housed in a dedicated, existing, octagonal top-end that allows for the complete assembly to be swapped during daytime operations.  The moving mass part of the ASM, is similar in depth to the current, solid secondary mirrors for WMKO (about 200 mm at the center). The envelope of the conceptual design allows for existing infrastructure on the telescope top-end, including the laser beam projector optics.

The assembly is mounted on a large hexapod structure that provides up to 50 mm of piston motion.  This focus range, along with the large stroke of the ASM, allows for aberration-free adjustment of the location of the Cassegrain focus by 1-2 m.  For example, movement of the Cassegrain focal plane 2 m further from its current location could be accomplished by moving the ASM 25.5 mm closer to the primary and applying a correction on the ASM for the spherical aberration introduced by this movement of about 5 $\mu$m peak-to-valley. The hexapod assembly also provides static tip, tilt and centration of the secondary, to sub-micron repeatability.   

The bright blue backing structure shown in the figure houses the actuators. The backing structure is silicon carbide, while the facesheet will likely be borofloat glass or similar material.  The coefficient of thermal expansion for these materials is quite similar, minimizing the potential distortion introduced by thermal changes.  The actuator electronics are located at the back end of the hexapod.  A kinematic interface exists between the ASM backing structure and the hexapod top ring to minimize deflections from hexapod movement. These locations also provide possible interface locations for fast tip-tilt actuators, should these be desired, to vibrationally isolate the ASM.  An initial finite element analysis of the structure indicates that resonant modes of the structure are at or above 750 Hz.  It is worth noting that, the actuators are designed to have resonant frequencies of around 1 kHz.  Since the actuators are rigidly connected to the facehseet the deformable shell has similarly high frequencies for its resonant modes.  This is different from voice coil-based ASM's with facesheet resonant modes that are much lower due to the lack of intrinsic actuator stiffness.  Such a design requires more complex internal control laws and co-located position sensors for high speed response and electronic damping\cite{Riccardi2003} of these low frequency shell vibrational modes.  

\begin{figure}[h!]
\centering
\begin{subfigure}{0.40\textwidth}
  \centering
  \includegraphics[width=1.0\linewidth]{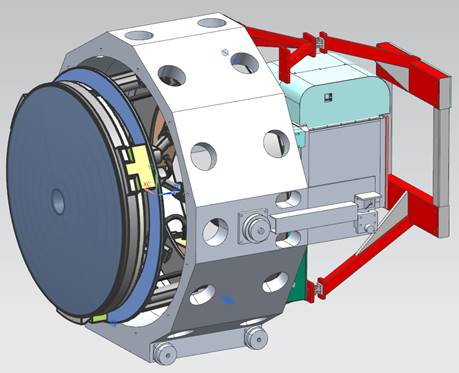}

\end{subfigure}
\begin{subfigure}{0.40\textwidth}
  \centering
   \includegraphics[width=1.0\linewidth]{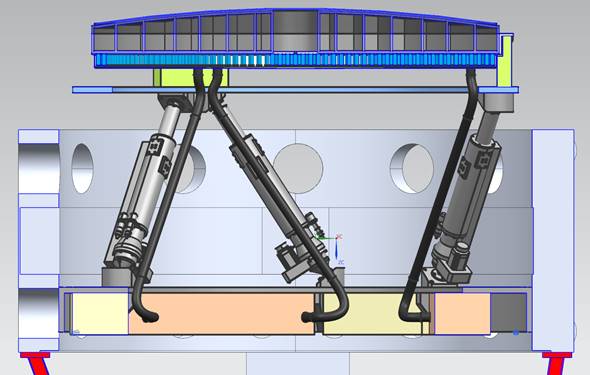}
 \end{subfigure}
 \caption[example]{Concept for an ASM for Keck.  The left image shows the device housed in an existing Keck top-end hexagonal structure.  The right image shows the device mounted on a hexapod capable for static alignment of the ASM.  
 }
   \end{figure}

\section{Conclusions}

ASM's that are simple, robust and high performance are being designed are around the hybrid variable reluctance actuators that have been recently developed by TNO.  Validation of this technology is underway through several prototype devices and the UH-88 ASM project. Here, we have presented initial concepts for ASM's that can be deployed on the APF and at WMKO.  The devices can be designed to work with and expand upon the current capabilities of these facilities.  The deployment of an ASM for the APF can provide improved efficiency and image stability, potentially leading to more efficient and stable radial velocity measurements.  An ASM for WMKO can provide both GLAO and diffraction-limited capabilities for a wider suite of instruments.  It has the potential to broaden the use of AO at WMKO, making adaptive correction of the wavefront available to any instruments capable of measuring the correction needed.  Development of a concept and strategy for leveraging this capability with current and future instruments will be important for defining a path forward for telescope-integrated AO at WMKO.

\bibliography{Adsec}   

\begin{thebibliography}{10}

\bibitem{Riccardi2010}
{Riccardi}, A., {Xompero}, M., {Briguglio}, R., {Quir{\'o}s-Pacheco}, F.,
  {Busoni}, L., {Fini}, L., {Puglisi}, A., {Esposito}, S., {Arcidiacono}, C.,
  {Pinna}, E., {Ranfagni}, P., {Salinari}, P., {Brusa}, G., {Demers}, R.,
  {Biasi}, R., and {Gallieni}, D., ``{The adaptive secondary mirror for the
  Large Binocular Telescope: optical acceptance test and preliminary on-sky
  commissioning results},'' in [{\em Adaptive Optics Systems
  II}{\nolinebreak\hspace{0.1em}]},  {Ellerbroek}, B.~L., {Hart}, M., {Hubin},
  N., and {Wizinowich}, P.~L., eds., {\em Society of Photo-Optical
  Instrumentation Engineers (SPIE) Conference Series} {\bf 7736},  77362C (July
  2010).

\bibitem{Biasi2010}
{Biasi}, R., {Gallieni}, D., {Salinari}, P., {Riccardi}, A., and {Mantegazza},
  P., ``{Contactless thin adaptive mirror technology: past, present, and
  future},'' in [{\em Adaptive Optics Systems II}{\nolinebreak\hspace{0.1em}]},
   {Ellerbroek}, B.~L., {Hart}, M., {Hubin}, N., and {Wizinowich}, P.~L., eds.,
  {\em Society of Photo-Optical Instrumentation Engineers (SPIE) Conference
  Series} {\bf 7736},  77362B (July 2010).

\bibitem{Kuiper2018}
{Kuiper}, S., {Doelman}, N., {Human}, J., {Saathof}, R., {Klop}, W., and
  {Maniscalco}, M., ``{Advances of TNO's electromagnetic deformable mirror
  development},'' in [{\em Advances in Optical and Mechanical Technologies for
  Telescopes and Instrumentation III}{\nolinebreak\hspace{0.1em}]},  {Navarro},
  R. and {Geyl}, R., eds., {\em Society of Photo-Optical Instrumentation
  Engineers (SPIE) Conference Series} {\bf 10706},  1070619 (July 2018).

\bibitem{Lloyd-Hart2000}
{Lloyd-Hart}, M., ``{Thermal Performance Enhancement of Adaptive Optics by Use
  of a Deformable Secondary Mirror},'' {\em pasp}~{\bf 112},  264--272 (Feb.
  2000).

\bibitem{Gillett1998}
{Gillett}, F. and {Mountain}, M., ``{On the Comparative Performance of an 8 M
  NGST and a Ground Based 8 M Optical/IR Telescope},'' in [{\em Science With
  The NGST}{\nolinebreak\hspace{0.1em}]},  {Smith}, E.~P. and {Koratkar}, A.,
  eds., {\em Astronomical Society of the Pacific Conference Series} {\bf 133},
  42 (Jan. 1998).

\bibitem{Masciadri2010}
{Masciadri}, E., {Stoesz}, J., {Hagelin}, S., and {Lascaux}, F., ``{Mt. Graham:
  optical turbulence vertical distribution with standard and high
  resolution},'' in [{\em Ground-based and Airborne Telescopes
  III}{\nolinebreak\hspace{0.1em}]},  {Stepp}, L.~M., {Gilmozzi}, R., and
  {Hall}, H.~J., eds., {\em Society of Photo-Optical Instrumentation Engineers
  (SPIE) Conference Series} {\bf 7733},  77331P (July 2010).

\bibitem{Avila2004}
{Avila}, R., {Masciadri}, E., {Vernin}, J., and {S{\'a}nchez}, L.~J.,
  ``{Generalized SCIDAR Measurements at San Pedro M{\'a}rtir. I. Turbulence
  Profile Statistics},'' {\em pasp}~{\bf 116},  682--692 (July 2004).

\bibitem{Osborn2018}
{Osborn}, J. and {Sarazin}, M., ``{Atmospheric turbulence forecasting with a
  general circulation model for Cerro Paranal},'' {\em mnras}~{\bf 480},
  1278--1299 (Oct. 2018).

\bibitem{Hagelin2008}
{Hagelin}, S., {Masciadri}, E., {Lascaux}, F., and {Stoesz}, J., ``{Comparison
  of the atmosphere above the South Pole, Dome C and Dome A: first attempt},''
  {\em mnras}~{\bf 387},  1499--1510 (July 2008).

\bibitem{Rabien2019}
{Rabien}, S., {Angel}, R., {Barl}, L., {Beckmann}, U., {Busoni}, L., {Belli},
  S., {Bonaglia}, M., {Borelli}, J., {Brynnel}, J., {Buschkamp}, P.,
  {Cardwell}, A., {Contursi}, A., {Connot}, C., {Davies}, R., {Deysenroth}, M.,
  {Durney}, O., {Eisenhauer}, F., {Elberich}, M., {Esposito}, S., {Frye}, B.,
  {Gaessler}, W., {Gasho}, V., {Gemperlein}, H., {Genzel}, R., {Georgiev},
  I.~Y., {Green}, R., {Hart}, M., {Kohlmann}, C., {Kulas}, M., {Lefebvre}, M.,
  {Mazzoni}, T., {Noenickx}, J., {Orban de Xivry}, G., {Ott}, T., {Peter}, D.,
  {Puglisi}, A., {Qin}, Y., {Quirrenbach}, A., {Raab}, W., {Rademacher}, M.,
  {Rahmer}, G., {Rosensteiner}, M., {Rix}, H.~W., {Salinari}, P., {Schwab}, C.,
  {Sivitilli}, A., {Steinmetz}, M., {Storm}, J., {Veillet}, C., {Weigelt}, G.,
  and {Ziegleder}, J., ``{ARGOS at the LBT. Binocular laser guided ground-layer
  adaptive optics},'' {\em aap}~{\bf 621},  A4 (Jan. 2019).

\bibitem{Chun2009}
{Chun}, M., {Wilson}, R., {Avila}, R., {Butterley}, T., {Aviles}, J.-L.,
  {Wier}, D., and {Benigni}, S., ``{Mauna Kea ground-layer characterization
  campaign},'' {\em mnras}~{\bf 394},  1121--1130 (Apr. 2009).

\bibitem{Chun2014}
{Chun}, M.~R., {Lai}, O., {Butterley}, T., {Goebel}, S., {Baranec}, C., and
  {Toomey}, D., ``{Extremely high-resolution ground-layer optical turbulence
  profile at Mauna Kea},'' in [{\em Adaptive Optics Systems
  IV}{\nolinebreak\hspace{0.1em}]},  {Marchetti}, E., {Close}, L.~M., and
  {Vran}, J.-P., eds., {\em Society of Photo-Optical Instrumentation Engineers
  (SPIE) Conference Series} {\bf 9148},  914867 (Aug. 2014).

\bibitem{Schok2009}
{Sch{\"o}ck}, M., {Els}, S., {Riddle}, R., {Skidmore}, W., {Travouillon}, T.,
  {Blum}, R., {Bustos}, E., {Chanan}, G., {Djorgovski}, S.~G., {Gillett}, P.,
  {Gregory}, B., {Nelson}, J., {Ot{\'a}rola}, A., {Seguel}, J., {Vasquez}, J.,
  {Walker}, A., {Walker}, D., and {Wang}, L., ``{Thirty Meter Telescope Site
  Testing I: Overview},'' {\em pasp}~{\bf 121},  384 (Apr. 2009).

\bibitem{Chun2018}
{Chun}, M., {Lu}, J., {Lai}, O., {Abdurrahman}, F., {Service}, M., {Toomey},
  D., {Fohring}, D., {Baranec}, C., {Hayano}, Y., and {Oya}, S., ``{On-sky
  results from the wide-field ground-layer adaptive optics demonstrator
  'imaka},'' in [{\em Adaptive Optics Systems VI}{\nolinebreak\hspace{0.1em}]},
   {Close}, L.~M., {Schreiber}, L., and {Schmidt}, D., eds., {\em Society of
  Photo-Optical Instrumentation Engineers (SPIE) Conference Series} {\bf
  10703},  107030J (July 2018).

\bibitem{Chun2010}
{Chun}, M.~R., {Carlberg}, R.~G., {Richer}, H.~B., {Mellier}, Y., {Astier}, P.,
  {Lai}, O., {Salmon}, D.~A., {Cuillandre}, J.-C., {Andersen}, D., {Pazder},
  J., {V{\'e}ran}, J.-P., {Barrick}, G.~A., {Bauman}, S., {Ho}, K.~K., {Avila},
  R., {Wilson}, R.~W., and {Butterley}, T., ``{'Imaka: a one-degree
  high-resolution imager for the Canada-France-Hawaii Telescope},'' in [{\em
  Ground-based and Airborne Instrumentation for Astronomy
  III}{\nolinebreak\hspace{0.1em}]},  {McLean}, I.~S., {Ramsay}, S.~K., and
  {Takami}, H., eds., {\em Society of Photo-Optical Instrumentation Engineers
  (SPIE) Conference Series} {\bf 7735},  77350I (July 2010).

\bibitem{Rigaut2018}
{Rigaut}, F., {Minowa}, Y., {Akiyama}, M., {Ono}, Y., {Korkiakoski}, V.,
  {Herrald}, N., {Gausachs}, G., {Clergeon}, C., {Wang}, S.-Y., {d'Orgeville},
  C., {Davies}, J., {Koyama}, Y., {Iwata}, I., {Kodama}, T., {Motohara}, K.,
  {Hayano}, Y., {Tanaka}, I., {Hattori}, T., and {Yoshida}, M., ``{A conceptual
  design study for Subaru ULTIMATE GLAO},'' in [{\em Adaptive Optics Systems
  VI}{\nolinebreak\hspace{0.1em}]},  {Close}, L.~M., {Schreiber}, L., and
  {Schmidt}, D., eds., {\em Society of Photo-Optical Instrumentation Engineers
  (SPIE) Conference Series} {\bf 10703},  1070324 (July 2018).

\bibitem{Lu2018}
{Lu}, J.~R., {Chun}, M., {Ammons}, S.~M., {Bundy}, K., {Dekany}, R., {Do}, T.,
  {Gavel}, D., {Kassis}, M., {Lai}, O., {Martin}, C.~L., {Max}, C., {Steidel},
  C., {Wang}, L., {Westfall}, K., and {Wizinowich}, P., ``{Ground layer
  adaptive optics for the W. M. Keck Observatory: feasibility study},'' in
  [{\em Adaptive Optics Systems VI}{\nolinebreak\hspace{0.1em}]},  {Close},
  L.~M., {Schreiber}, L., and {Schmidt}, D., eds., {\em Society of
  Photo-Optical Instrumentation Engineers (SPIE) Conference Series} {\bf
  10703},  107030N (July 2018).

\bibitem{Szeto2006}
{Szeto}, K., {Andersen}, D., {Crampton}, D., {Morris}, S., {Lloyd-Hart}, M.,
  {Myers}, R., {Jensen}, J.~B., {Fletcher}, M., {Gardhouse}, W.~R., {Milton},
  N.~M., {Pazder}, J., {Stoesz}, J., {Simons}, D., and {V{\'e}ran}, J.-P., ``{A
  proposed implementation of a ground layer adaptive optics system on the
  Gemini Telescope},'' in [{\em Society of Photo-Optical Instrumentation
  Engineers (SPIE) Conference Series}{\nolinebreak\hspace{0.1em}]},  {McLean},
  I.~S. and {Iye}, M., eds., {\em Society of Photo-Optical Instrumentation
  Engineers (SPIE) Conference Series} {\bf 6269},  626958 (June 2006).

\bibitem{Kuiper2020}
{Kuiper}, S. and {et al.}, ``{Performance Analysis of an adaptive secondary
  mirror for the University of Hawaii 2.2-meter Telescope},'' in [{\em Adaptive
  Optics Systems VII}{\nolinebreak\hspace{0.1em}]},  {\em Society of
  Photo-Optical Instrumentation Engineers (SPIE) Conference Series} {\bf 11448}
  (2020).

\bibitem{Jonker2020}
{Jonker}, W.~A. and {et al.}, ``{Design and manufacturing status of the UH-88
  adaptive secondary mirror},'' in [{\em Adaptive Optics Systems
  VII}{\nolinebreak\hspace{0.1em}]},  {\em Society of Photo-Optical
  Instrumentation Engineers (SPIE) Conference Series} {\bf 11451} (2020).

\bibitem{Chun2020}
{Chun}, M.~R., {Baranec}, C., {Lai}, O., {Lu}, J., {Zhang}, S., {Kuiper}, S.,
  {Jonker}, W., and {Maniscalco}, M., ``{A new adaptive secondary mirror for
  astronomy on the University of Hawaii 2.2-meter Telescope},'' in [{\em
  Adaptive Optics Systems VII}{\nolinebreak\hspace{0.1em}]},  {\em Society of
  Photo-Optical Instrumentation Engineers (SPIE) Conference Series} {\bf 11448}
  (2020).

\bibitem{Bowens-Rubin2020}
{Bowens-Rubin}, R. and et~al., ``{Performance of Large-Format Deformable
  Mirrors Constructed with TNO Variable Reluctance Actuators},'' in [{\em
  Adaptive Optics Systems VII}{\nolinebreak\hspace{0.1em}]},  {\em Society of
  Photo-Optical Instrumentation Engineers (SPIE) Conference Series} {\bf 11448}
  (2020).

\bibitem{Por2018}
{Por}, E.~H., {Haffert}, S.~Y., {Radhakrishnan}, V.~M., {Doelman}, D.~S., {van
  Kooten}, M., and {Bos}, S.~P., ``{High Contrast Imaging for Python (HCIPy):
  an open-source adaptive optics and coronagraph simulator},'' in [{\em
  Adaptive Optics Systems VI}{\nolinebreak\hspace{0.1em}]},  {Close}, L.~M.,
  {Schreiber}, L., and {Schmidt}, D., eds., {\em Society of Photo-Optical
  Instrumentation Engineers (SPIE) Conference Series} {\bf 10703},  1070342
  (July 2018).

\bibitem{LoCurto2010}
{Lo Curto}, G., {Lovis}, C., {Wilken}, T., {Avila}, G., {Chazelas}, B.,
  {Esposito}, M., {H{\"a}nsch}, T.~W., {Gonz{\'a}ez-Hern{\'a}ndez}, J.,
  {Holzwarth}, R., {Ihle}, G., {Manescau}, A., {Pasquini}, L., {Pepe}, F.,
  {Rebolo}, R., {Segovia}, A., {Sinclaire}, P., {Steinmetz}, T., {Udem}, T.,
  and {Wildi}, F., ``{Along the path towards extremely precise radial velocity
  measurements},'' in [{\em Ground-based and Airborne Instrumentation for
  Astronomy III}{\nolinebreak\hspace{0.1em}]},  {McLean}, I.~S., {Ramsay},
  S.~K., and {Takami}, H., eds., {\em Society of Photo-Optical Instrumentation
  Engineers (SPIE) Conference Series} {\bf 7735},  77350Z (July 2010).

\bibitem{Crane2010}
{Crane}, J.~D., {Shectman}, S.~A., {Butler}, R.~P., {Thompson}, I.~B., {Birk},
  C., {Jones}, P., and {Burley}, G.~S., ``{The Carnegie Planet Finder
  Spectrograph: integration and commissioning},'' in [{\em Ground-based and
  Airborne Instrumentation for Astronomy III}{\nolinebreak\hspace{0.1em}]},
  {McLean}, I.~S., {Ramsay}, S.~K., and {Takami}, H., eds., {\em Society of
  Photo-Optical Instrumentation Engineers (SPIE) Conference Series} {\bf 7735},
   773553 (July 2010).

\bibitem{Burt2014}
{Burt}, J., {Vogt}, S.~S., {Butler}, R.~P., {Hanson}, R., {Meschiari}, S.,
  {Rivera}, E.~J., {Henry}, G.~W., and {Laughlin}, G., ``{The Lick-Carnegie
  Exoplanet Survey: Gliese 687 b{\textemdash}A Neptune-mass Planet Orbiting a
  Nearby Red Dwarf},'' {\em apj}~{\bf 789},  114 (July 2014).

\bibitem{Vogt2014}
{Vogt}, S.~S., {Radovan}, M., {Kibrick}, R., {Butler}, R.~P., {Alcott}, B.,
  {Allen}, S., {Arriagada}, P., {Bolte}, M., {Burt}, J., {Cabak}, J.,
  {Chloros}, K., {Cowley}, D., {Deich}, W., {Dupraw}, B., {Earthman}, W.,
  {Epps}, H., {Faber}, S., {Fischer}, D., {Gates}, E., {Hilyard}, D., {Holden},
  B., {Johnston}, K., {Keiser}, S., {Kanto}, D., {Katsuki}, M., {Laiterman},
  L., {Lanclos}, K., {Laughlin}, G., {Lewis}, J., {Lockwood}, C., {Lynam}, P.,
  {Marcy}, G., {McLean}, M., {Miller}, J., {Misch}, T., {Peck}, M., {Pfister},
  T., {Phillips}, A., {Rivera}, E., {Sandford}, D., {Saylor}, M., {Stover}, R.,
  {Thompson}, M., {Walp}, B., {Ward}, J., {Wareham}, J., {Wei}, M., and
  {Wright}, C., ``{APF{\textemdash}The Lick Observatory Automated Planet
  Finder},'' {\em pasp}~{\bf 126},  359 (Apr. 2014).

\bibitem{Wizinowich2006}
{Wizinowich}, P.~L., {Le Mignant}, D., {Bouchez}, A.~H., {Campbell}, R.~D.,
  {Chin}, J. C.~Y., {Contos}, A.~R., {van Dam}, M.~A., {Hartman}, S.~K.,
  {Johansson}, E.~M., {Lafon}, R.~E., {Lewis}, H., {Stomski}, P.~J., {Summers},
  D.~M., {Brown}, C.~G., {Danforth}, P.~M., {Max}, C.~E., and {Pennington},
  D.~M., ``{The W. M. Keck Observatory Laser Guide Star Adaptive Optics System:
  Overview},'' {\em pasp}~{\bf 118},  297--309 (Feb. 2006).

\bibitem{Park2014}
{Park}, C., {Jaffe}, D.~T., {Yuk}, I.-S., {Chun}, M.-Y., {Pak}, S., {Kim},
  K.-M., {Pavel}, M., {Lee}, H., {Oh}, H., {Jeong}, U., {Sim}, C.~K., {Lee},
  H.-I., {Nguyen Le}, H.~A., {Strubhar}, J., {Gully-Santiago}, M., {Oh}, J.~S.,
  {Cha}, S.-M., {Moon}, B., {Park}, K., {Brooks}, C., {Ko}, K., {Han}, J.-Y.,
  {Nah}, J., {Hill}, P.~C., {Lee}, S., {Barnes}, S., {Yu}, Y.~S., {Kaplan}, K.,
  {Mace}, G., {Kim}, H., {Lee}, J.-J., {Hwang}, N., and {Park}, B.-G.,
  ``{Design and early performance of IGRINS (Immersion Grating Infrared
  Spectrometer)},'' in [{\em Ground-based and Airborne Instrumentation for
  Astronomy V}{\nolinebreak\hspace{0.1em}]},  {Ramsay}, S.~K., {McLean}, I.~S.,
  and {Takami}, H., eds., {\em Society of Photo-Optical Instrumentation
  Engineers (SPIE) Conference Series} {\bf 9147},  91471D (July 2014).

\bibitem{Riccardi2003}
{Riccardi}, A., {Brusa}, G., {Salinari}, P., {Gallieni}, D., {Biasi}, R.,
  {Andrighettoni}, M., and {Martin}, H.~M., ``{Adaptive secondary mirrors for
  the Large Binocular Telescope},'' in [{\em Adaptive Optical System
  Technologies II}{\nolinebreak\hspace{0.1em}]},  {Wizinowich}, P.~L. and
  {Bonaccini}, D., eds., {\em Society of Photo-Optical Instrumentation
  Engineers (SPIE) Conference Series} {\bf 4839},  721--732 (Feb. 2003).

\end{thebibliography}
\bibliographystyle{spiebib}   

\end{document}